\documentclass[a4paper,11pt]{article}
\usepackage{pos}

\title{
\vskip-3cm{\baselineskip14pt
    \begin{flushright}
     \normalsize \normalfont{TTP24-023, P3H-24-045, LTH-1377, ZU-TH 33/24}
    \end{flushright}} \vskip2.5cm
    
Electroweak corrections to $gg \to HH$: Factorizable
contributions}

\author[a]{Joshua Davies}
\author[b]{Kay Sch\"onwald}
\author[c]{Matthias Steinhauser}
\author*[c]{Hantian Zhang}

\affiliation[a]{Department of Mathematical Sciences, University of Liverpool, \\Liverpool, L69 3BX, UK}

\affiliation[b]{Physik-Institut, Universit\"at Z\"urich, \\
Winterthurerstrasse 190, 8057 Z\"urich, Switzerland}

\affiliation[c]{Institut f{\"u}r Theoretische Teilchenphysik, Karlsruhe Institute of Technology (KIT), \\
Wolfgang-Gaede Strasse 1, 76128 Karlsruhe, Germany}

\abstract{
In these proceedings, we consider the factorizable contributions of the two-loop electroweak corrections to Higgs boson pair production at the Large Hadron Collider.
We discuss the classification of factorizable diagrams and their renormalization.
Compact analytic results for the renormalized form factors are presented in a computer-readable form.
}

\FullConference{Loops and Legs in Quantum Field Theory (LL2024)\\
 14-19, April, 2024\\
Wittenberg, Germany\\}

\begin{document}
\maketitle


\section{Introduction}

Higgs boson pair production will play an important role in constraining the Higgs self-coupling  in the
coming years at the Large Hadron Collider (LHC) at CERN.  Thus, besides
the QCD corrections also the higher order electroweak corrections are
important.

The complete electroweak (EW) NLO corrections to $gg\to HH$ have been computed in
Ref.~\cite{Bi:2023bnq}  although in a completely numerical approach.  
A different numerical approach has been applied for the top-Yukawa and Higgs self-coupling corrections in Ref.~\cite{Heinrich:2024dnz}.
For practical purposes it is often useful to have also analytic results,
even if they are only valid in a restricted region of phase space. Up
to now a few first steps have been undertaken in this direction.
As a first well-defined subclass all contributions involving four
Higgs-top Yukawa couplings have been considered in Ref~\cite{Davies:2022ram}. A
deep expansion has been computed in the high-energy limit and it
has been shown that precise results can be obtained once Pad\'e
approximants are constructed, even close to the top quark pair
threshold.

The leading $m_t^4$ and $m_t^2$ corrections have been considered in
Ref.~\cite{Muhlleitner:2022ijf}.  These results have been confirmed in Ref.~\cite{Davies:2023npk}
and extended by considering all Feynman diagrams involving a top
quark. An expansion up to $1/m_t^{10}$ has been computed in the
large-$m_t$ limit. In this reference also the renormalization is
discussed taking into account all sectors of the Standard Model.

\begin{figure}[hbt!]
  \begin{tabular}{cccc}
    \includegraphics[width=.25\textwidth]{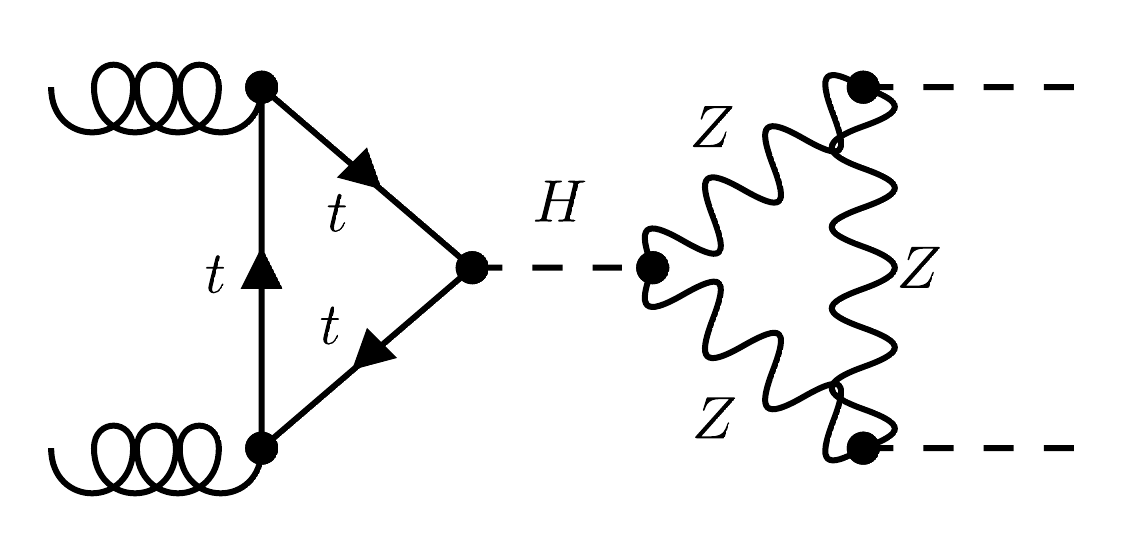} &
    \includegraphics[width=.24\textwidth]{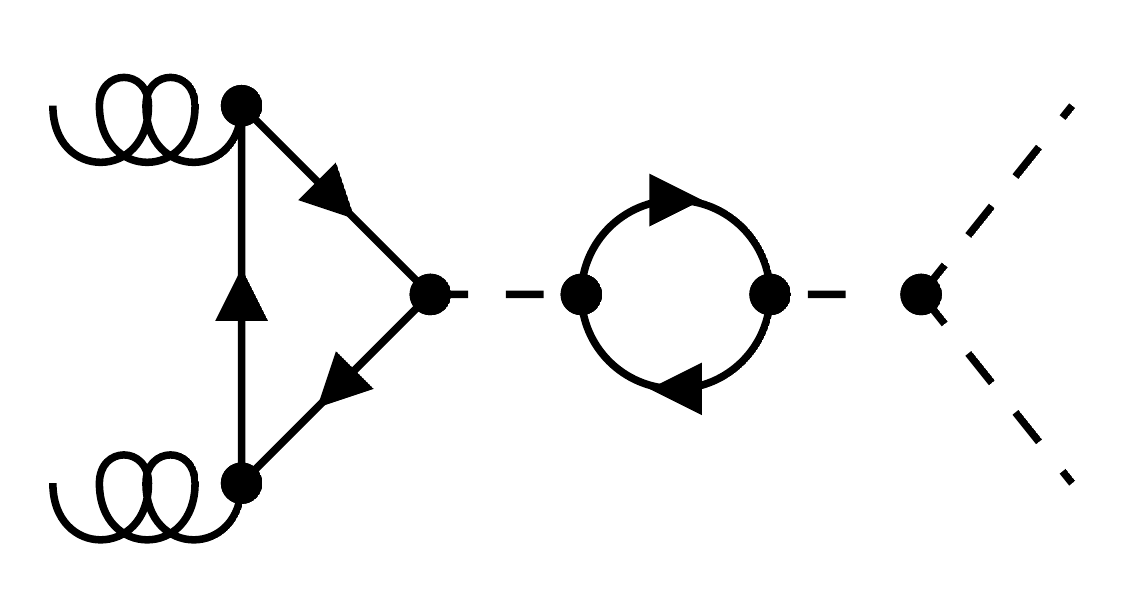} &
    \includegraphics[width=.2\textwidth]{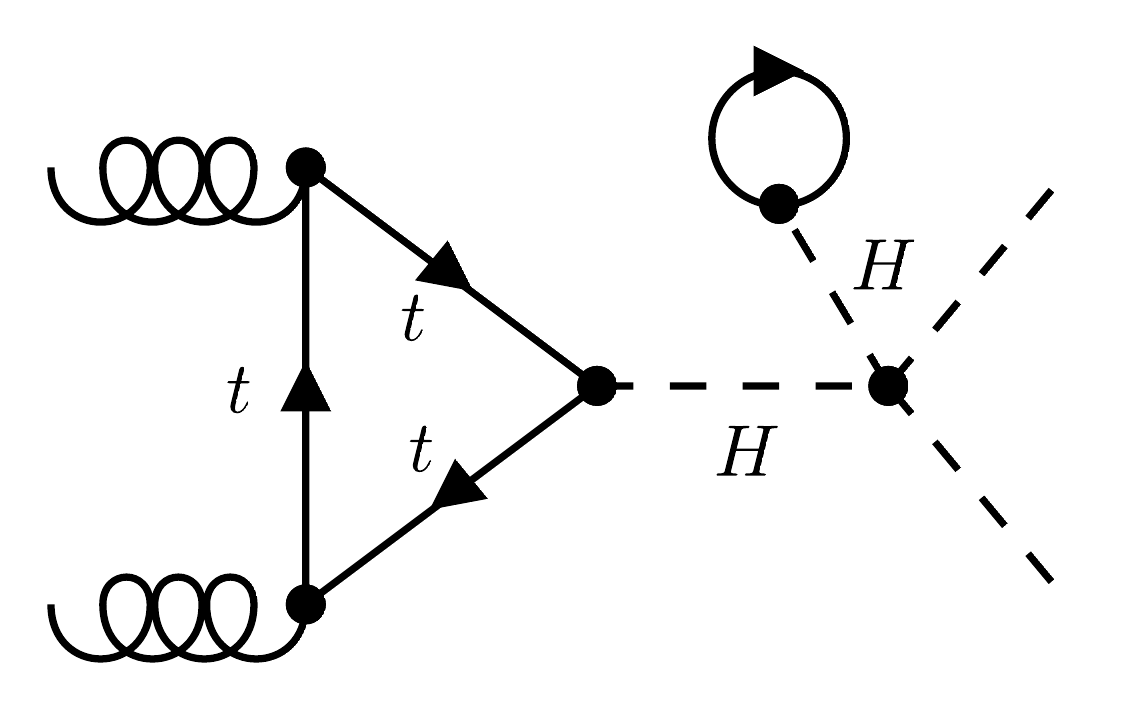} &
    \includegraphics[width=.2\textwidth]{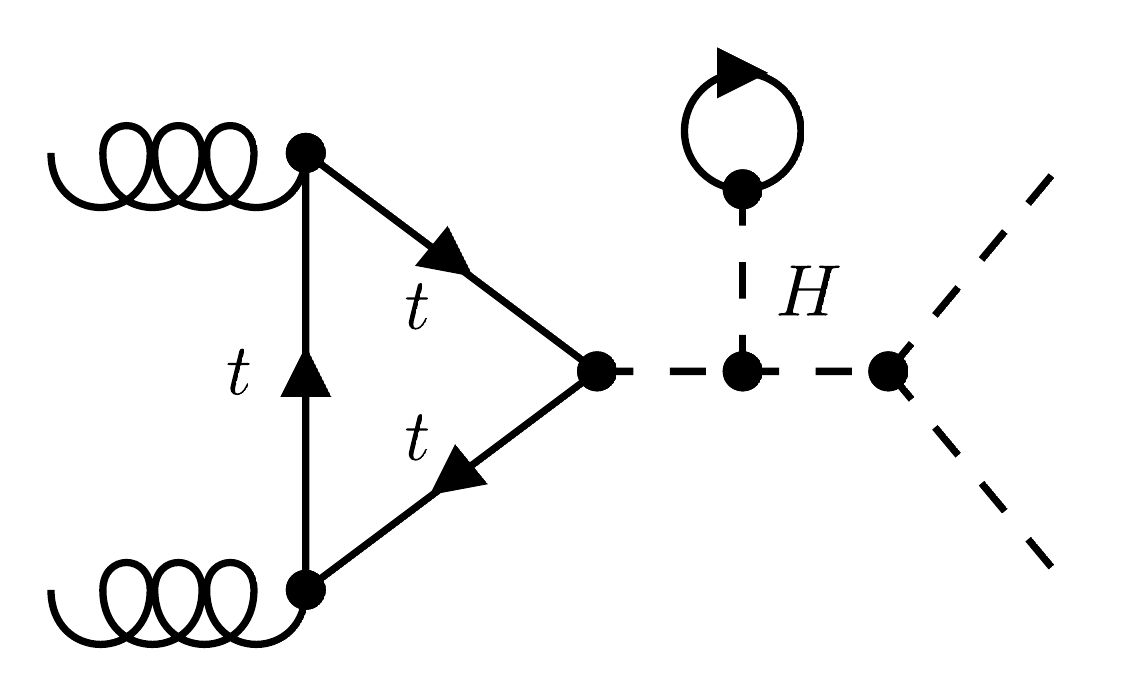} \\
    class 1 & class 2 & class 3 & class 4 \\
    \includegraphics[width=.22\textwidth]{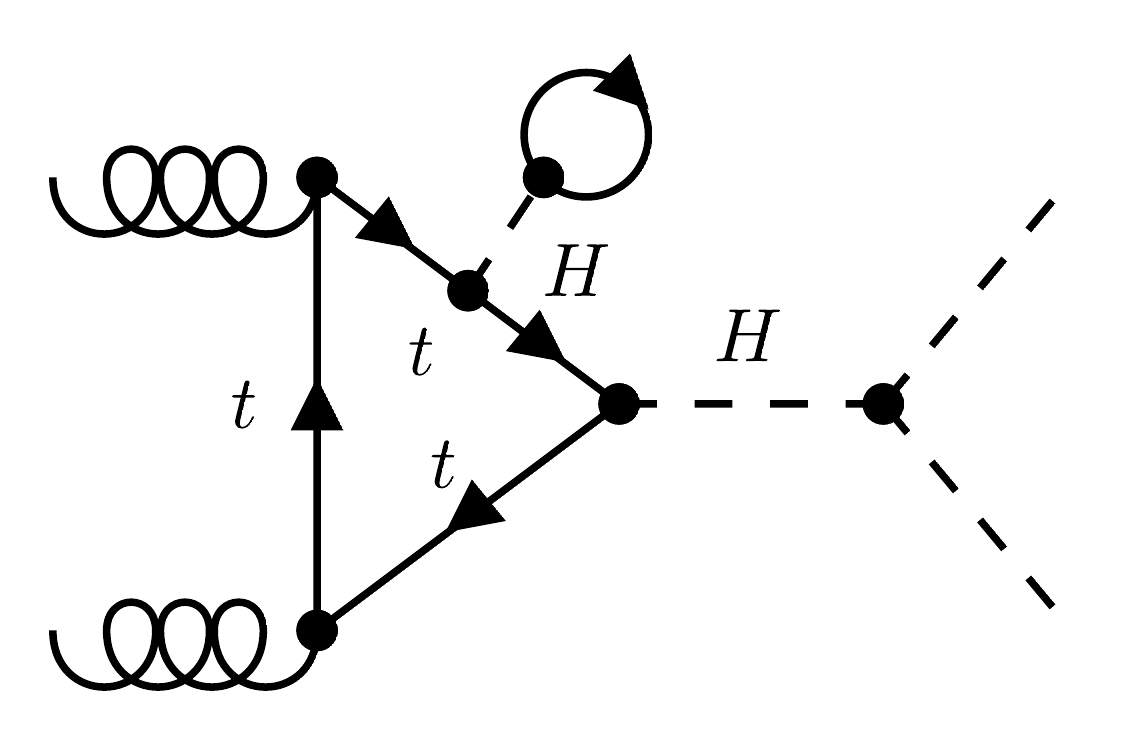} &
    \includegraphics[width=.22\textwidth]{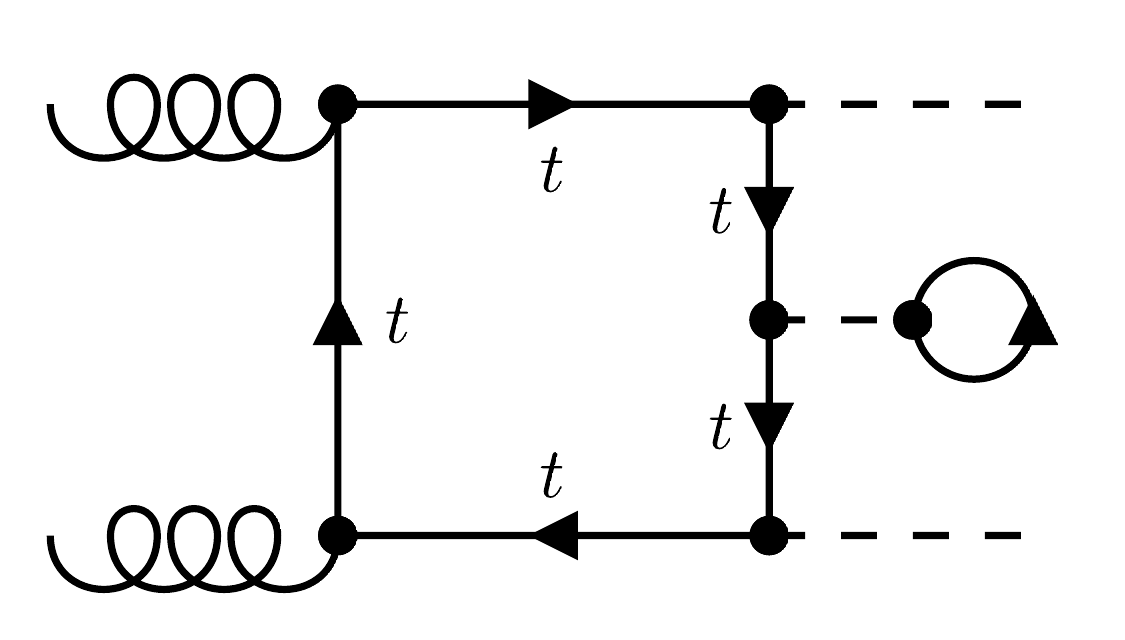} & & \\
    class 5 & class 6 &  & 
  \end{tabular}
  \caption{\label{fig::diags}
   Classification of  Feynman diagrams contributing to factorizable contributions.
    The six diagrams represent the six classes
    of diagrams which we distinguish in this paper.}
\end{figure}

In this contribution we compute exact analytic results for the factorizable contribution to Higgs boson pair production in gluon fusion.
Sample Feynman diagram are shown in Fig.~\ref{fig::diags}.
In the next section we provide details regarding the calculation,
Section~\ref{sec::ren} deals with the renormalization
and Section~\ref{sec::res} discusses the results and
provides a summary.

\section{Technical details}

For the generation of the Feynman amplitudes we use {\tt qgraf}~\cite{Nogueira:1991ex} and
process them afterwards with {\tt tapir}~\cite{Gerlach:2022qnc} and {\tt exp}~\cite{Harlander:1998cmq,Seidensticker:1999bb}.
We use the {\tt FORM}-based~\cite{Ruijl:2017dtg} setup {\tt calc} for the manipulation of
the $\gamma$ matrix algebra, the application of projectors, and the reduction
to master integrals.  In the following we provide more details regarding some of
these elements.

The amplitude for the process $g(q_1)g(q_2)\to H(q_3)H(q_4)$ (with
$q_4=-q_1-q_2-q_3$) is conveniently decomposed into two Lorentz structures
\begin{eqnarray}
  {\cal M}^{ab} &=& 
  \varepsilon_{1,\mu}\varepsilon_{2,\nu}
  {\cal M}^{\mu\nu,ab}
  \,\,=\,\,
  \varepsilon_{1,\mu}\varepsilon_{2,\nu}
  \delta^{ab} X_0 s 
  \left( F_1 A_1^{\mu\nu} + F_2 A_2^{\mu\nu} \right)
  \,,
                    \label{eq::M}
\end{eqnarray}
where $a$ and $b$ are adjoint colour indices  and the 
Lorentz structures are given by
\begin{eqnarray}
  A_1^{\mu\nu} &=& g^{\mu\nu} - {\frac{1}{q_{12}}q_1^\nu q_2^\mu
  }\,,\nonumber\\
  A_2^{\mu\nu} &=& g^{\mu\nu}
                   + \frac{1}{{p_T^2} q_{12}}\left(
                   q_{33}    q_1^\nu q_2^\mu
                   - 2q_{23} q_1^\nu q_3^\mu
                   - 2q_{13} q_3^\nu q_2^\mu
                   + 2q_{12} q_3^\mu q_3^\nu \right)\,.
\end{eqnarray}
In this equation the following abbreviations have been introduced
\begin{eqnarray}
  q_{ij} &=& q_i\cdot q_j\,,\qquad
  {p_T^{\:2}} \:\:\:=\:\:\: \frac{2q_{13}q_{23}}{q_{12}}-q_{33}
  {\:\:\:=\:\:\: \frac{tu - m_H^4}{s}} \,,
\end{eqnarray}
where $s=(q_1+q_2)^2$ is the
squared partonic centre-of-mass energy, and ${t=(q_1+q_3)^2}$, ${u=(q_2+q_3)^2}$ are Mandelstam variables
which fulfill $s+t+u=2m_H^2$, and the prefactor $X_0$ is given by
\begin{eqnarray}
  X_0 &=& \frac{G_F}{\sqrt{2}} \frac{\alpha_s(\mu)}{2\pi} T_F \,,
\end{eqnarray}
with $T_F=1/2$, $G_F$ is Fermi's constant and $\alpha_s(\mu)$ is the strong
coupling constant evaluated at the renormalization scale $\mu$.
We define perturbative expansion of the form factors as
\begin{align}
F_i & = F_i^{(0)} + \frac{\alpha}{\pi} F_i^{(1)} + \dots
\end{align}
where $\alpha$ is the fine structure constant.

Equation~(\ref{eq::M}) defines the two form factors
$F_1$ and $F_2$. 
Note that the Feynman diagrams considered in this
paper  from class 1 to 4 in Fig.~\ref{fig::diags} only contribute to $F_1$ since $F_2$ only gets contributions
from box diagrams.
These diagrams all have at least one bridge connecting the initial and final states.
In practice, only Higgs bosons serve as bridges.
Note that diagrams with $Z$ bosons or neutral Goldstone bosons do not contribute,
as the vector coupling part vanishes due to the charge conjugation symmetry, 
and the axial coupling part vanishes due to the parity-even properties of the process.
We will provide final results for the renormalized form factor $F_1$ in  Feynman gauge.
Note that the finite part of the renormalized factorizable form  factor $F_1$ depends on $\xi_{W}$ and $\xi_Z$.
They will drop out once also the non-factorizable contribution is added. 
This has been checked in Ref.~\cite{Davies:2023npk} in the large-$m_t$ limit.

We classify the factorizable contribution into six groups which are shown in
Fig.~\ref{fig::diags}. Classes 5 and 6 contribute to the
renormalization of the top quark mass and they cancel exactly if the top quark
is renormalized on shell.\footnote{Note that box diagrams
  in class~6 provide contributions both to $F_1$ and $F_2$.} In our calculation the corresponding diagrams are
included, however, to avoid confusion, we discard them in the final expressions
provided in the ancillary file.
The remaining diagrams can be classified according to
the number of $s$-channel Higgs boson propagators, $D_h(s)$
and  propagators without momentum flow, $D_h(0)$.
The latter arise from diagrams with tadpoles.
\begin{itemize}
  \item class 1: single $D_h(s)$, no $D_h(0)$
  \item class 2: double $D_h(s)$, no $D_h(0)$
  \item class 3: single $D_h(s)$ and single $D_h(0)$
  \item class 4: double $D_h(s)$ and single $D_h(0)$
\end{itemize}
As we will discuss in Section~\ref{sec::ren} these
classes renormalize separately.

For the practical calculation one only has to consider one-loop
integrals which are well known in the literature. However, it is important that in intermediate
steps   higher order $\epsilon$ terms are properly
taken into account. Thus, in our calculation we refrain from using published packages since they are either targeted to the limit $d \to 4$ or the handling of higher order $\epsilon$ terms is not transparent.
Rather, we define box-type integral
families within our \texttt{FORM} setup with external momenta $q_1^2=q_2^2=0$ and $q_3^2\not=0\not=q_4^2$.
In our routines we even have $q_3^2\not=q_4^2$ although for the current
application $q_3^2=q_4^2=m_H^2$ is sufficient.  We use {\tt
  LiteRed}~\cite{Lee:2012cn} to generate generic reduction rules which we then
convert to {\tt FORM} code. This allows us to perform the reduction to master
integrals within the {\tt calc} setup in a fast and efficient way for general
dimensional parameter $\epsilon$.

After inserting explicit result for the one-loop tadpole contributions we can
express our final result in terms of (products of) one-loop two- and
three-point functions. More precisely, we have the $\epsilon^0$ term of
two-point functions with squared external momentum $s$, $t$ or $m_H^2$. Both
internal lines have the same mass, either $m_t, m_W, m_Z$ or $m_H$.
Furthermore, there are two types of three-point functions. Both of them have
the virtuality $s$ on one of the external legs and the other two are both
either zero or $m_H^2$.  All three lines have again the same mass.  The bare
expression also contains the term of order $\epsilon$ of the three-point
function with external momenta $\{q_1^2, q_2^2, (q_1+q_2)^2\} = \{0,0,s\}$ and internal top quark mass. As we will
see in Section~\ref{sec::ren}, this contribution cancels after
renormalization.


In our final results, we introduce the following notation for these functions
\begin{align}
& B_0^{\rm fin} \big( q^2, m_X^2, m_X^2 \big) \; :  \;\; \mbox{$\epsilon^0$-part of two-point function} \nonumber \\
& C_0 \big( q_1^2, q_2^2, (q_1+q_2)^2, m_X^2, m_X^2, m_X^2 \big)  \; :  \;\;  \mbox{$\epsilon^0$-part of three-point function} \nonumber  \\
& C_0^{\epsilon} \big( q_1^2, q_2^2, (q_1+q_2)^2, m_X^2, m_X^2, m_X^2 \big)  \; :  \;\;  \mbox{$\epsilon^1$-part of three-point function}
\label{eq:C0epDef}
\end{align}
where $m_X$ is the internal particle mass.
These notations are quite standard and we have adopted them from \texttt{LoopTools}~\cite{Hahn:1998yk}.
To clarify our convention and normalization, we provide the two-point and three-point functions 
including $\mathcal{O}(\epsilon)$ terms
\begin{align}
B_0^{\rm fin}(s, m_t^2, m_t^2) \;=\; &  \log \left(\frac{\mu^2}{m_t^2}\right)+\frac{\sqrt{s (s-4 m_t^2)} \log (x)}{s}+2 \nonumber \\
C_0 \big( 0,0,s,m_t^2,m_t^2,m_t^2 \big) \; = \; & \frac{\log(x)^2}{2 s}  \,, \nonumber \\
C_0^{\epsilon} \big( 0,0,s,m_t^2,m_t^2,m_t^2 \big) \; = \; &      
 \frac{\log ^2(x) \log \left(\frac{\mu^2}{m_t^2}\right)}{2 s}-\frac{4 \text{Li}_3(-x)}{s} -\log (x) \left(\frac{\pi ^2}{6 s}-\frac{2 \text{Li}_2(-x)}{s}\right) \nonumber \\
 & +\frac{\log ^3(x)}{6 s}-\frac{3 \zeta (3)}{s} \,, 
 \label{eq:C0ep}
\end{align} 
where $x = \big(\sqrt{s (s-4 m_t^2)}+2 m_t^2-s \big)/(2 m_t^2)$.
As presented, the formula is valid for $s>4m_t^2$. It can be evaluated in other kinematical regions by 
proper analytic continuation taking into account $s \to s+ \mathrm{i} \, \delta$.




\section{\label{sec::ren}Renormalization}

The renormalization of the complete contribution (i.e. factorizable
and non-factorizable) is discussed in Ref.~\cite{Davies:2023npk} where
it has successfully been applied in the large-$m_t$ limit.  From the
considerations in~\cite{Davies:2023npk} it is straightforward to extract the
following prescription which leads to a finite results for the factorizable contribution:
\begin{itemize}
\item At LO we only consider the triangle diagram and renormalize the
  parameters $ \{\lambda,v,m_H\}$ associated to the Higgs propagator and trilinear coupling in the on-shell scheme.  
Here $\lambda$ is the Higgs boson self-coupling, $v$ is the vacuum expectation value, $m_H$ is the Higgs boson mass.
  Note that the top-Yukawa coupling $y_t$ and top-mass $m_t$ are not renormalized.
  
\item In addition to the parameter renormalization we have to take into
  account the wave function renormalization of the external Higgs boson
  field. For the factorizable part half of this contribution is needed; the other half
  is needed for the non-factorizable contribution.
\end{itemize}

In practice, we use the relation
\begin{align}
\lambda & = \frac{m_H^2}{2 v^2} \,, \quad v = \frac{2 m_W \sin \theta_{\rm W}}{e}
\end{align}
where $\theta_{\rm W}$ is the weak mixing angle and $\cos \theta_{\rm W} = m_W /m_Z$.
We use this relation between bare and renormalized parameters for $e,m_H,m_W,m_Z$ where $e$ is the electric charge.
Note that in all parts we retain all tadpole contributions from Higgs boson
lines except those where the Higgs boson is attached to a top quark line.
The latter contribution is closely related to the top quark mass
renormalization in the non-factorizable contribution (see also discussion above).

The main results of this contribution are obtained in Feynman gauge. In
principle, a calculation for generic gauge parameters ($\xi_W$ and $\xi_Z$) is
possible but tedious since in that case different masses appear inside the
loop functions.  In Ref.~\cite{Davies:2023npk} we have incorporated the $\xi_W$
and $\xi_Z$ terms in the limits $\xi_{W}, \xi_Z \gg 1$ and have checked that in the
sum of all contributions the gauge parameters drop out. In other kinematic
limits we can proceed in a similar way.



\section{\label{sec::res}Results and Summary}
We present our final results in  electronic form~\cite{ttplink}.
We introduce labels for each counterterm contribution such that 
the bare result is obtained if they are set to zero.
Furthermore, there are labels for the four classes 1 to 4.
%
Note that the $C_0^{\epsilon}$ introduced in Eq.\eqref{eq:C0epDef}  drops out after renormalization.
For reference we provide in the following the one-loop triangle form factor up to $\mathcal{O}(\epsilon)$
\begin{align}
F_1^{(0)} \;=\; & -\frac{6 m_h^2 m_t^2 \log ^2(x) (s-4 m_t^2)}{s^2  (s-m_h^2 )}+\frac{24 m_h^2 m_t^2}{s  (s-m_h^2 )}  
+ \epsilon \, \Bigg[\frac{48 m_h^2 m_t^2 \text{Li}_3(-x) (s-4 m_t^2)}{s^2  (s-m_h^2 )} \nonumber \\
& -\frac{24 m_h^2 m_t^2 \text{Li}_2(-x) \log (x)  (s-4 m_t^2 )}{s^2  (s-m_h^2 )}+\log ^2(x)  \Bigg(\frac{24 m_h^2 m_t^4}{s^2  (s-m_h^2 )}-\frac{6 m_h^2 m_t^2  (s-4 m_t^2 ) \log  \Big(\frac{\mu ^2}{m_t^2} \Big)}{s^2  (s-m_h^2 )} \Bigg) \nonumber \\
& -\frac{2 m_h^2 m_t^2   (s-4 m_t^2 ) \log ^3(x)}{s^2  (s-m_h^2 )}+\frac{2 m_h^2 m_t^2   \Big(12 \sqrt{s  (s-4 m_t^2 )}+\pi ^2 (s - 4m_t^2)  \Big) \log (x)}{s^2  (s-m_h^2 )} \nonumber \\
& +\frac{36 \zeta (3) m_h^2 m_t^2  (s-4 m_t^2 )}{s^2  (s-m_h^2 )}+\frac{24 m_h^2 m_t^2 \log  \Big(\frac{\mu ^2}{m_t^2} \Big)}{s  (s-m_h^2 )}+\frac{72 m_h^2 m_t^2}{s  (s-m_h^2 )}\Bigg] \,,
\end{align}
where $x$ is defined after Eq.\eqref{eq:C0ep}.
Note that
our final expression only contains scalar functions and no tensor coefficient functions.
The  analytic results are well known and easy to evaluate numerically
with packages like \texttt{LoopTools}~\cite{Hahn:1998yk}, \texttt{PackageX}~\cite{Patel:2016fam}, \texttt{Collier}~\cite{Denner:2016kdg}, \texttt{OneLoop}~\cite{vanHameren:2010cp} and other packages.
For a subset of our results involving top-Yukawa and Higgs self-coupling contributions, 
we have compared to Ref.~\cite{Heinrich:2024dnz} and found agreement.

To summarize, in these proceedings we present a further building block to the
NLO electroweak corrections to the process $gg\to HH$: the complete factorizable
contributions.  We provide our results in a computer-readable file
which makes an implementeation in a numerical code
straightforward.


\section{Acknowledgements}
This research was supported by the Deutsche Forschungsgemeinschaft (DFG, German Research Foundation) under grant 396021762 — TRR 257 “Particle Physics Phenomenology after the Higgs Discovery” and has received funding from the European Research Council (ERC) under the European Union’s Horizon 2020 research and innovation programme grant agreement 101019620 (ERC Advanced Grant TOPUP). The work of JD is supported by the STFC Consolidated Grant ST/X000699/1.
The authors thank G. Heinrich, S. P. Jones, M. Kerner, T. W. Stone and A. Vestner  for sharing their unpublished results for comparisons.

\begin{thebibliography}{10}

\bibitem{Bi:2023bnq}
H.-Y. Bi, L.-H. Huang, R.-J. Huang, Y.-Q. Ma and H.-M. Yu, \emph{{Electroweak
  Corrections to Double Higgs Production at the LHC}},
  \href{https://doi.org/10.1103/PhysRevLett.132.231802}{\emph{Phys. Rev. Lett.}
  {\bfseries 132} (2024) 231802}
  [\href{https://arxiv.org/abs/2311.16963}{{\ttfamily 2311.16963}}].

\bibitem{Heinrich:2024dnz}
G.~Heinrich, S. P.~Jones, M.~Kerner, T. W.~Stone and A.~Vestner
\emph{Electroweak corrections to Higgs boson pair production: The top-Yukawa and self-coupling contributions}, 
[\href{https://arxiv.org/abs/2407.04653}{{\ttfamily
  2407.04653}}].

\bibitem{Davies:2022ram}
J.~Davies, G.~Mishima, K.~Sch\"onwald, M.~Steinhauser and H.~Zhang,
  \emph{{Higgs boson contribution to the leading two-loop Yukawa corrections to
  gg \textrightarrow{} HH}},
  \href{https://doi.org/10.1007/JHEP08(2022)259}{\emph{JHEP} {\bfseries 08}
  (2022) 259} [\href{https://arxiv.org/abs/2207.02587}{{\ttfamily
  2207.02587}}].

\bibitem{Muhlleitner:2022ijf}
M.~M\"uhlleitner, J.~Schlenk and M.~Spira, \emph{{Top-Yukawa-induced
  corrections to Higgs pair production}},
  \href{https://doi.org/10.1007/JHEP10(2022)185}{\emph{JHEP} {\bfseries 10}
  (2022) 185} [\href{https://arxiv.org/abs/2207.02524}{{\ttfamily
  2207.02524}}].

\bibitem{Davies:2023npk}
J.~Davies, K.~Sch\"onwald, M.~Steinhauser and H.~Zhang, \emph{{Next-to-leading
  order electroweak corrections to $gg \to HH$ and $gg \to gH$ in the
  large-$m_t$ limit}},
  \href{https://doi.org/10.1007/JHEP10(2023)033}{\emph{JHEP} {\bfseries 10}
  (2023) 033} [\href{https://arxiv.org/abs/2308.01355}{{\ttfamily
  2308.01355}}].

\bibitem{Nogueira:1991ex}
P.~Nogueira, \emph{{Automatic Feynman Graph Generation}},
  \href{https://doi.org/10.1006/jcph.1993.1074}{\emph{J. Comput. Phys.}
  {\bfseries 105} (1993) 279}.

\bibitem{Gerlach:2022qnc}
M.~Gerlach, F.~Herren and M.~Lang, \emph{{tapir: A tool for topologies,
  amplitudes, partial fraction decomposition and input for reductions}},
  \href{https://doi.org/10.1016/j.cpc.2022.108544}{\emph{Comput. Phys. Commun.}
  {\bfseries 282} (2023) 108544}
  [\href{https://arxiv.org/abs/2201.05618}{{\ttfamily 2201.05618}}].

\bibitem{Harlander:1998cmq}
R.~Harlander, T.~Seidensticker and M.~Steinhauser, \emph{{Complete corrections
  of Order alpha alpha-s to the decay of the Z boson into bottom quarks}},
  \href{https://doi.org/10.1016/S0370-2693(98)00220-2}{\emph{Phys. Lett. B}
  {\bfseries 426} (1998) 125}
  [\href{https://arxiv.org/abs/hep-ph/9712228}{{\ttfamily hep-ph/9712228}}].

\bibitem{Seidensticker:1999bb}
T.~Seidensticker, \emph{{Automatic application of successive asymptotic
  expansions of Feynman diagrams}},  in \emph{{6th International Workshop on
  New Computing Techniques in Physics Research: Software Engineering,
  Artificial Intelligence Neural Nets, Genetic Algorithms, Symbolic Algebra,
  Automatic Calculation}}, 5, 1999,
  \href{https://arxiv.org/abs/hep-ph/9905298}{{\ttfamily hep-ph/9905298}}.

\bibitem{Ruijl:2017dtg}
B.~Ruijl, T.~Ueda and J.~Vermaseren, \emph{{FORM version 4.2}},
  \href{https://arxiv.org/abs/1707.06453}{{\ttfamily 1707.06453}}.

\bibitem{Lee:2012cn}
R.~N. Lee, \emph{{Presenting LiteRed: a tool for the Loop InTEgrals
  REDuction}},  \href{https://arxiv.org/abs/1212.2685}{{\ttfamily 1212.2685}}.

\bibitem{ttplink}
\verb|https://www.ttp.kit.edu/preprints/2024/ttp24-023/|.

\bibitem{Hahn:1998yk}
T.~Hahn and M.~Perez-Victoria, \emph{{Automatized one loop calculations in
  four-dimensions and D-dimensions}},
  \href{https://doi.org/10.1016/S0010-4655(98)00173-8}{\emph{Comput. Phys.
  Commun.} {\bfseries 118} (1999) 153}
  [\href{https://arxiv.org/abs/hep-ph/9807565}{{\ttfamily hep-ph/9807565}}].

\bibitem{Patel:2016fam}
H.~H. Patel, \emph{{Package-X 2.0: A Mathematica package for the analytic
  calculation of one-loop integrals}},
  \href{https://doi.org/10.1016/j.cpc.2017.04.015}{\emph{Comput. Phys. Commun.}
  {\bfseries 218} (2017) 66}
  [\href{https://arxiv.org/abs/1612.00009}{{\ttfamily 1612.00009}}].

\bibitem{Denner:2016kdg}
A.~Denner, S.~Dittmaier and L.~Hofer, \emph{{Collier: a fortran-based Complex
  One-Loop LIbrary in Extended Regularizations}},
  \href{https://doi.org/10.1016/j.cpc.2016.10.013}{\emph{Comput. Phys. Commun.}
  {\bfseries 212} (2017) 220}
  [\href{https://arxiv.org/abs/1604.06792}{{\ttfamily 1604.06792}}].

\bibitem{vanHameren:2010cp}
A.~van Hameren, \emph{{OneLOop: For the evaluation of one-loop scalar
  functions}}, \href{https://doi.org/10.1016/j.cpc.2011.06.011}{\emph{Comput.
  Phys. Commun.} {\bfseries 182} (2011) 2427}
  [\href{https://arxiv.org/abs/1007.4716}{{\ttfamily 1007.4716}}].

\end{thebibliography}

\providecommand{\href}[2]{#2}\begingroup\raggedright\endgroup

\end{document}